\documentclass[12pt]{iopart}

\usepackage{iopams}
\usepackage{graphicx}
\usepackage{setspace}
\usepackage[dvips,usenames]{color}
\usepackage{fancyhdr}
\scrollmode

\newcommand{\ie}{{\it i.e.}}
\begin{document}

\fancyhf{}
\fancyhead[L]{\textit{\nouppercase{Distribution of Schmidt-like eigenvalues}}}
\fancyhead[R]{\nouppercase{M. P. Pato and G. Oshanin }}
\fancyfoot[C]{\thepage}

\title[Distribution of Schmidt-like eigenvalues]{Distribution of Schmidt-like
eigenvalues for Gaussian Ensembles of the Random Matrix Theory}


\author{Mauricio P. Pato$^{1}$ and Gleb Oshanin$^{2}$}

\address{$^{1}$Instituto de F\'{i}sica, Universidade de S\~{a}o Paulo,
C.P. 66318, 05314-970 S\~{a}o Paulo, S.P., Brazil}
\address{$^{2}$Laboratoire de Physique Th\'{e}orique de la Mati\`{e}re
Condens\'{e}e (UMR CNRS 7600),\\
Universit\'{e} Pierre et Marie Curie, 4 Place Jussieu, 75252 Paris Cedex
5 France}

\eads{\mailto{mpato@fma.if.usp.br}, \mailto{oshanin@lptmc.jussieu.fr}}

\begin{abstract}

We analyze the form of the probability
distribution function $P_{n}^{( \beta)}(w)$
of the Schmidt-like random variable
$w = x_1^2/\left(\sum_{j=1}^n
    x^{2}_j/n\right)$, where $x_j$ are the eigenvalues of a given $n \times n$ $\beta$-Gaussian
    random matrix, $\beta$ being the Dyson symmetry index. This variable, by definition, can be
considered as a measure
of how any individual eigenvalue deviates from
the arithmetic mean value of all eigenvalues of
a given random matrix, and its distribution is 
calculated with respect to the ensemble of such $\beta$-Gaussian
    random matrices. 
We show that in the asymptotic limit $n \to \infty$ and for arbitrary $\beta$ the distribution  $P_{n}^{( \beta)}(w) $ converges to the  Mar\v{c}enko-Pastur form, i.e., is defined as $P_{n}^{( \beta)}(w) \sim \sqrt{(4 - w)/w}$ for $w \in [0,4]$ and equals zero outside of the support.
Furthermore, for Gaussian unitary ($\beta = 2$) ensembles 
we present exact explicit expressions
for $P_{n}^{( \beta=2)}(w)$ which are valid for arbitrary $n$
 and
analyze their behavior.
\end{abstract}

\vspace{2pc}

\vspace{2pc}
\journal{ JSTAT}
\maketitle

\section{Introduction}

Random covariance matrices were introduced by
J. Wishart  in his studies of
multivariate populations \cite{Wishart}.
In physical literature, statistical  properties of the eigenvalues of random matrices
have attracted a great deal of attention 
since the seminal works of Wigner \cite{wigner},
Dyson \cite{dyson,dyson1} and Mehta \cite{Mehta}. Various random variables associated with eigenvalues of
random matrices have been analyzed,
such as, e.g., gaps in the eigenvalue spectra,
number of eigenvalues in a given interval, largest or smallest eigenvalues and etc,
with a special
emphasis
put on their typical or atypical behavior. A variety of results
and their relevance to physical systems have been recently discussed in Ref.~\cite{boh,satya}.

One of such variables
is the so-called Schmidt eigenvalue,
used to characterize, e.g., the degree of entanglement of random pure states in bipartite
quantum systems. It is
defined as one of the eigenvalues of a given random matrix
divided by the trace, \ie, the sum of all eigenvalues.
On physical grounds, this variable can be therefore
considered as a measure of heterogeneity 
of the eigenvalues and shows
how any individual eigenvalue deviates from
the arithmetic mean of all eigenvalues of 
a given random matrix.
A number of significant results on the
distributions of such eigenvalues and their extreme values
for $\beta$-Laguerre-Wishart matrices
have been obtained (see, e.g., Refs.\cite{1,2,3,4,11} and references therein).
Such random variables
have also been considered recently
within a different context as probes of an effective broadness of the first passage time
distributions in bounded domains \cite{5,6}.

In this paper we analyze the forms of the probability
distribution function $P_{n}^{( \beta)}(w)$
of a Schmidt-like random variable
\begin{equation}
\label{def}
w = \frac{x^{2}_{1}}{\sum_{j=1}^n
    x^{2}_j/n} \,,
\end{equation}
where $x_j$ are the eigenvalues of a given $n \times n$ $\beta$-Gaussian random matrix and $\beta$ is
the Dyson symmetry index.  Note that we use a term "Schmidt-like random variable" since here 
we define $w$ as the ratio of a \textit{squared} eigenvalue over the sum
of all \textit{squared} eigenvalues.
Within such a definition $w$
is always positive definite and has a support on $[0,n]$.

The probability distribution function $P_{n}^{(\beta)}(w)$ is given by
\begin{equation}
P_{n}^{( \beta)}(w)=\left< \delta \left(w- \frac{nx^{2}_{1}}{\sum
    x^{2}_j}\right)\right> ,   \label{1}
\end{equation}
where the average is to be calculated with the weight 
\begin{equation}
P(x_1,x_2,...,x_n) = \frac{1}{K_n}\exp\left(-\frac{\beta}{2}
\sum_{k=1}^{n} x_{k}^2\right)\prod_{j>i}|x_j - x_i |^{\beta} \,,
\end{equation}
$K_n$ being a known normalization constant \cite{Mehta}.
Our aim is to determine an asymptotic behavior of 
$P_{n}^{( \beta)}(w)$ 
for arbitrary $\beta$ and $n \to \infty$. Apart of this, we will present an exact, explicit results for Gaussian Unitary ($\beta=2$)  Ensembles (GUE) valid    
for arbitrary $n$.

The paper is outlined as follows: In section \ref{general} we provide some general results for
$P_{n}^{(  \beta)}(w)$ and analyze its asymptotic forms when $n \to \infty$. In section \ref{unitary}  we present explicit results for the distribution function of the Schmidt-like eigenvalues
for GUE. Finally, in section \ref{conclusion} we conclude with a brief recapitulation of our results.

\section{Asymptotic behavior for arbitrary $\beta$}
\label{general}

Taking advantage of the Fourier cosine representation of the delta-function,
we can conveniently rewrite Eq.~(\ref{1}) as

\begin{equation}
P_{n}^{( \beta)}(w)=\frac{2}{\pi}\int_0 ^{\infty}dy\cos(w y)
\left< \cos\left( \frac{n y x^{2}_{1}}{\sum    x^{2}_j}\right)\right> \,,
\end{equation}
so that, expanding the cosine into the Taylor series, we obtain

\begin{equation}
\fl
P_{n}^{( \beta)}(w)=\frac{2}{\pi}\int_0 ^{\infty}dy\cos(w y) \int_{-\infty}^{\infty} dx \, \rho_{n}^{( \beta)} (x)
\sum_{k=0}^{\infty} \frac{ (-1)^{k}}{(2k)!}
\left< \left(\frac{nyx^{2}_{1}}{\sum x^{2}_j}\right)^{2k} \right> \,,
\end{equation}
where $\rho_{n}^{( \beta)} (x)$ is the eigenvalue density.
Next, using the integral identity

\begin{equation}
\left(\sum
  x^{2}_j\right)^{-2k}=\frac{1}{\Gamma(2k)}\int_{0}^{\infty}d\xi \, \exp\left(-\xi\sum
  x^{2}_j\right) \xi^{2k-1} \,,
\end{equation}
we can straightforwardly calculate the multiple integrals over $d x_i$ with $i=2,3,...n$, which yields

\begin{equation}
\fl
\label{hh}
P_{n}^{( \beta)}(w)=\frac{2}{\pi}\int_0 ^{\infty}dy \, \cos(w y) \int_{-\infty}^{\infty} dx \rho_{n}^{( \beta)} (x)
\sum_{k=0}^{\infty} (-1)^{k} \frac{\Gamma(f_{\beta}/2)}{(2k)!\Gamma(2k+f_{\beta}/2)}
\left(\frac{\beta nyx^{2}}{2}\right)^{2k}   ,
\end{equation}
where
$f_{\beta}=n+\beta n(n-1)/2$.
Further on, performing the summation over $k$ in Eq. (\ref{hh}), we obtain

\begin{equation}
\fl
P_{n}^{( \beta)}(w)=\frac{\Gamma(f_{\beta}/2)}{\pi} \, \int_0^{\infty} dy \, \cos(w y) \int_{-\infty}^{\infty} dx \, \rho_{n}^{(\beta)} (x)
\left(\mbox{\bf I}_{f_{\beta}/2 -1} \left(x\sqrt{2\beta i n y}\right) +
cc\right)   \label{8}
\end{equation}
where $ \mbox{\bf I}_{\nu} (z)=(z/2)^{-\nu}\mbox{I}_{\nu} (z)    $,
$\mbox{I}_{\nu} (z)$ being the modified Bessel function, and "cc" stands for the complex conjugate of
$ \mbox{\bf I}_{\nu} (z)$.
Equation (\ref{8}) constitutes our main general result valid for arbitrary $\beta$.

The result in Eq. (\ref{8}) allows us to establish,  for arbitrary $\beta$,  the
limiting asymptotic behavior  of the distribution when $n \to \infty$.
To do this, we first replace the function $ \mbox{\bf I}_{\nu} (z)$ by its integral
representation \cite{Abram}

\begin{equation}
\mbox{\bf I}_{\nu} (z) = \frac{1}{\Gamma(\nu +1/2)\sqrt{\pi}}\int_{-1}^{1} dt
\exp(zt) \left(1-t^2\right)^{\nu-1/2}   \,,  \label{9}
\end{equation}
and notice that for large values of $n$ the exponential part in the integrand in Eq.~(\ref{9}) is a strongly oscillating function
of the argument $z$. This permits us to make the replacement
$(1-t^2)^{\nu - 1/2} \approx \exp\left(-\nu t^2\right)$,
such that after the substitution
$u=t\sqrt{\nu},$ Eq.~(\ref{9}) becomes

\begin{equation}
\fl
\mbox{\bf I}_{\nu} (z) = \frac{\exp(-\frac{z^2}{4\nu})}
{\Gamma(\nu+1/2)\sqrt{\nu \pi}}\int_{-\sqrt{\nu}}^{\sqrt{\nu}} du
\exp\left(-(u-\frac{z}{2\sqrt{\nu}})^2\right)   \approx
\frac{1}{\Gamma(\nu+1/2)\sqrt{\nu }}\exp\left(-\frac{z^2}{4\nu}\right) \,.
\end{equation}

Substituting the latter equation into Eq.~(\ref{8}), we arrive at the following
representation

\begin{equation}
\fl
P_{n}^{( \beta)}(w)=\frac{2\Gamma(f_{\beta}/2)}{\pi\sqrt{f_{\beta}/2}\Gamma[(f_{\beta}-1)/2]}
\int_0 ^{\infty}dy\cos(w y)\int_{-\infty}^{\infty} dx \, \rho_{n}^{( \beta)} (x)
\cos \left(\frac{x^2\beta ny}{f_{\beta}}\right) \,,  \label{8a}
\end{equation}
which yields, after performing the integrations, the following asymptotic form

\begin{equation}
P_{n}^{( \beta)}(w)= \sqrt{\frac{n}{2w}} \, \rho_{n}^{(\beta)}\left(\sqrt{\frac{nw}{2}}\right) \,,
\end{equation}
i.e., it simply expresses  the desired probability distribution of the Schmidt-like
random variable $w$  through the eigenvalue density with an appropriately rescaled variable. The asymptotic behavior of the latter is well-known and is defined by the Wigner semi-circle distribution  \cite{wigner,Mehta}, so that after some very straightforward calculations we find
the following asymptotic form for the normalized probability distribution function :
\begin{equation}
P_{\infty}^{( \beta)}(w)= \frac{1}{2\pi}
\left\{
\begin{array}{rl}
\sqrt{\frac{4-w}{w}} , \mbox{  for  } 0 < w < 4 \,, \\
0 ,  \mbox{   for  } w > 4 \,.
\end{array}
\right.
\label{sc}
\end{equation}
Equation (\ref{sc}) holds 
for any value of the Dyson symmetry index $\beta$. 
It might seem surprising at the first glance 
that the limiting distribution  in Eq.~(\ref{sc}) has the form
of the Mar\v{c}enko-Pastur law \cite{mar}.  On the other hand, recall that as $n \to \infty$,
 the eigenvalues tend to be equidistantly-spaced so that the sum $\sum_j x_j^2/n$ tends to a constant. Then, it becomes clear
 why the distribution  $P_{n}^{( \beta)}(w)$ converges to an appropriately normalized single eigenvalue density, defined by the semi-circle distribution
  \cite{wigner,Mehta}, so that its squared value is distributed according to the Mar\v{c}enko-Pastur law.

\section{Gaussian Unitary Ensemble}
\label{unitary}

We turn now to the GUE case ($\beta = 2$ and $f_2 = n^2$), aiming to evaluate an explicit expression
 for the probability distribution $P_{n}^{( 2)}(w)$, valid for an arbitrary value of $n$. In this case,
 the eigenvalue density is given by

\begin{equation}
\rho_{n}^{(2)} (x)= \frac{\exp(-x^2)}{2^{n}n!\sqrt{\pi}}\left[H_n ^2
    (x)-H_{n+1}(x)H_{n-1}(x)\right] \,,   \label{10}
\end{equation}
where $H_n(x)$ denotes the Hermite polynomial \cite{Abram}.
Using Eqs.~(\ref{9}) and (\ref{10}), we can represent
the  integral as

\begin{equation}
\fl
 \int_{-\infty}^{\infty}dx \, \rho_{n}^{(2)}(x) \, \mbox{\bf I}_{f_2/2 -1}\left(2 x\sqrt{iny}\right) =
\int_{-1}^{1} \frac{dt
  \left(1-t^2\right)^{(n^{2}-3)/2}e^{inyt^2}}{2^{n}n!
\Gamma(n^{2} /2 -1/2)\pi} f(t\sqrt{iny}) \, ,
\end{equation}
where

\begin{equation}
f(t\sqrt{iny})=\int_{-\infty}^{\infty}dxe^{-(x-t\sqrt{iny})^2}  \left[H_n ^2
(x)    -H_{n+1}(x)H_{n-1}(x)\right]  .
\end{equation}
Further on, this function can be expressed in terms of the associated Laguerre
polynomials $L_n^{(\alpha)}(p)$ since for $n\geq m$ \cite{Abram}

\begin{equation}
\int_{-\infty}^{\infty}dxe^{-(x-z)^2}H_m(x) H_n(x)= 2^n \sqrt{\pi}m!z^{n-m}L_{m}^{(n-m)}(-2z^2) \,,
\end{equation}
 which leads to

\begin{equation}
f(t\sqrt{iny})=2^n \sqrt{\pi} L_{n-1}^{(1)} (-2t^2 iny) ,
\end{equation}
where we made use of the following recurrence relation between the associated Laguerre polynomials : $ nL_n (p)+ pL_{n-1}^{(2)}(p)=L_{n-1}^{(1)} (p)$.
Then, the probability distribution function $P_{n}^{( 2)}(w)$ becomes

\begin{equation}
\fl
P_{n}^{(2)}(w)=C
\int_{0}^{\infty}dy\cos(w y)\int_{0}^{1} \frac{dv}{\sqrt{v}}\left(1-v\right)^{(n^{2}-3)/2}
\left(e^{i n y v} L_{n-1}^{(1)} (-2  i n y v)  +   cc\right) \,,
\end{equation}
where the normalization constant $C$ is given explicitly by
\begin{equation}
C=\frac{\pi^{-3/2}\Gamma(n^{2}/2)}{n\Gamma[(n^{2}-1)/2]} \,,
\end{equation}
and $"cc"$ here stands for the complex conjugate of $\exp\left(i n y v\right) L_{n-1}^{(1)} (-2 i n y v) $.
Using next the series representation of the Laguerre polynomials,
the integration over $d y$ reduces to the calculation of the following integrals

\begin{equation}
\sum_{0}^{n-1}a_{k} \int_{0}^{\infty} dy\cos(w y) \left[
e^{inyv}  (-2v iny)^k +  e^{-inyt^2}  (2v iny)^k \right] \,,
\end{equation}
or

\begin{equation}
2\sum_{0}^{n-1}a_{k} \int_{0}^{\infty} dy\cos(w y)
\cos\left(nyv+\frac{k\pi}{2}\right)(-2vny)^k \,,
\end{equation}
which, using the fact that $\cos(x+k\pi/2)=\frac{d^k}{dx^k}\cos x,$
can be put into the following form

\begin{equation}
\pi L_{n-1}^{(1)} \left(-2v\frac{d}{dv} \right) \delta(w-nv) \,.
\end{equation}
Next, the integration over $d v$ can be performed by parts, taking into account that
$v^{k-1/2}(1-v)^{(n^{2}-3)/2}$ with $k>0$ vanishes
at the integration limits. Then, the operator expression

\begin{equation}
P_{n}^{(2)}(w)=\frac{\Gamma(n^2/2)}{\sqrt{\pi}n^2\Gamma[(n^2-1)/2]}\left.
 L_{n-1}^{(1)} \left(2\frac{d}{dv}v\right)
\frac{\left(1-v\right)^{(n^2-3)/2}}{\sqrt{v}}
\right|_{v=w/n}
\end{equation}
is obtained where powers of the  polynomial operator are understood
as  $2^j\frac{d^{j}}{dv^{j}}v^{j}.$ Explicitly, the action of the Laguerre polynomial
operator is defined as

\begin{equation}
\fl
 L_{n-1}^{(1)} \left(2\frac{d}{dv}v\right) \frac{\left(1-v\right)^{(n^2-3)/2}}{\sqrt{v}} = \sum_{k=0}^{n-1}
\frac{(-2)^k n!}{(n-1-k)!(k+1)!k!}\frac{d^k}{dv^k} \left(v^{k-1/2}
\left(1-v\right)^{(n^2-3)/2}\right) ,
\end{equation}
where the derivatives can be identified with the Rodrigues formula
for the Jacobi polynomials $P_n^{(a,b)}(p)$ \cite{Abram} as

\begin{eqnarray}
\fl
\frac{d^k}{dv^k} \left(v^{k-1/2} \left(1-v\right)^{(n^2-3)/2}\right) =
  \frac{(-1)^k k!\left(1-v\right)^{(n^2-3)/2-k}}{\sqrt{v}} \,
P_k^{((n^2-3)/2-k,-1/2)} (2v-1) \,.
\end{eqnarray}
Consequently, recalling that $v = w/n$,
we find
the following explicit 
result for the distribution $P_{n}^{(2)}(w)$ :

\begin{eqnarray}
\label{main}
\fl
P_{n}^{(2)}(w) &=& \frac{\Gamma(n^2/2)}{\sqrt{\pi} \, \Gamma\left((n^2-1)/2\right) n^{3/2}} \, \frac{\left(1 -w/n\right)^{(n^2-2 n - 1)/2}}{ \sqrt{w}} \,\times\nonumber\\
\fl
&\times& \sum_{k=0}^{n-1} {n \choose k + 1} \,
2^k \, \left(1 - \frac{w}{n}\right)^{n-k-1} \, P_{k}^{((n^2-3)/2-k,-1/2)} \left(2 \, \frac{w}{n} - 1\right)
  \,.  \end{eqnarray}
Further on, the sum on the right-hand-side
 of the latter equation
 can be also represented, after some straightforward calculations, as a polynomial
 of $w$, which yields
 \begin{eqnarray}
\label{main2}
\fl
P_{n}^{(2)}(w) &=& \frac{2 \, \Gamma(n^2/2)}{\pi^{3/4} \, \Gamma\left((n^2-1)/2\right) n^{1/2}} \, \frac{\left(1 -w/n\right)^{(n^2-2 n - 1)/2}}{ \sqrt{w}} \, \times \nonumber\\
 \fl
 &\times& \sum_{m=0}^{n-1} \frac{(-1)^m}{n^m} {n - 1 \choose m} \, \alpha_m \, w^m
  \,,
 \end{eqnarray}
 where the coefficients $\alpha_m$ are defined by
 \begin{equation}
\alpha_m = \int^1_0 \frac{\sqrt{1-t} \, dt}{\sqrt{t}} \, \left(1 - 2 t\right)^{n - m - 1} \, _2F_1\left(-m, \frac{n^2}{2} - 1,\frac{1}{2},2 t\right) \,,
 \end{equation}
 $_2F_1\left(\cdot\right)$ being the Gauss hypergeometric function. Equations (\ref{main}) and (\ref{main2}) constitute our principal results for case of the Gaussian Unitary Ensemble.

Before we turn to the analysis of the
asymptotic behavior of the distribution function, it
might be expedient
to present several first $P_{n}^{(2)}(w)$ explicitly.
Below we display $P_{n}^{(2)}(w)$ for $n = 2, 3, 4, 5$ and $6$:

\begin{equation}
\fl
\label{two}
P_{2}^{(2)}(w)= \frac{1}{\pi}\frac{1}{\sqrt{(2-w) \, w}}  \,,
\end{equation}

\begin{equation}
\fl
P_{3}^{(2)}(w) = \frac{35 \, (3 - w)}{576 \sqrt{3} \sqrt{w}} \, \Big(3 - 2 \, w + 3 \, w^2\Big) \,,\end{equation}

\begin{equation}
\fl
P_{4}^{(2)}(w) = \frac{(4 - w)^{7/2}}{1716 \, \pi \, \sqrt{w}} \, \Big(12 + 30 \, w - 53 \, w^2 + 38 \, w^3\Big) \,,
\end{equation}

\begin{equation}
\fl
P_{5}^{(2)}(w) = \frac{2028117 \, (5- w)^7}{81920000000000 \, \sqrt{5} \, \sqrt{w}} \, \Big(375 -300 \, w + 4490 \, w^2 -5996 \, w^3 +2711 \, w^4\Big) \,,
\end{equation}
and

\begin{eqnarray}
\fl
\label{six}
P_{6}^{(2)}(w) = \frac{32768  \, (6 - w)^{23/2}}{25113523969051155 \, \pi \, \sqrt{w}} \, \Big(810 &+& 3780 \, w - 18090 \, w^2 \nonumber\\
\fl
&+& 52878 \, w^3 - 49567 \, w^4 + 16144 \, w^5\Big) \,.
\end{eqnarray}

\begin{figure}[ht]
   \centerline{\includegraphics*[width=0.65\textwidth]{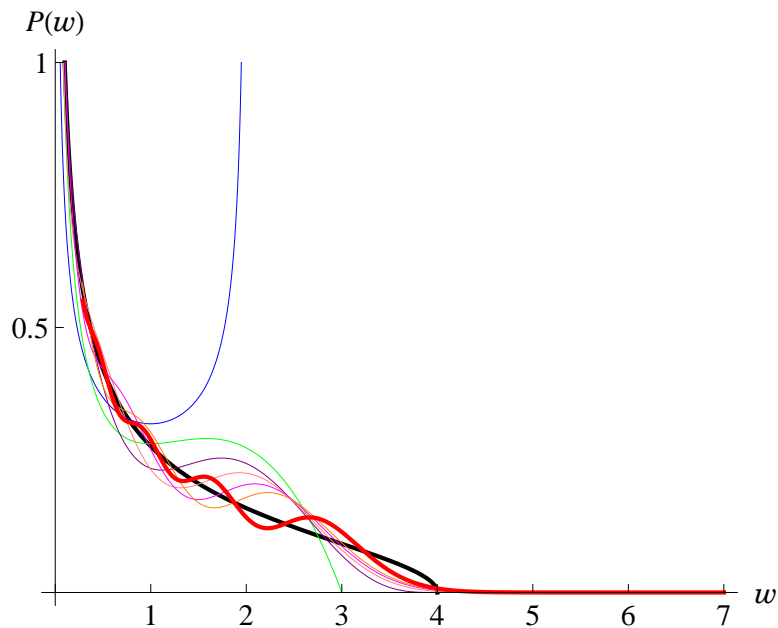}}
   \caption{(color online) The distribution $P_n^{(2)}(w)$ in Eqs.~(\ref{main}) and (\ref{main2})
   for $n = 2$ (blue), $n=3$ (green), $n=4$ (purple), $n=5$ (pink), $n=6$ (magenta), $n=7$ (orange) and $n=12$ (red). Solid black line defines the asymptotic result in Eq.~(\ref{sc}).}
\label{fig0}
\end{figure}

One notices that the expressions in Eqs.(\ref{two}) to (\ref{six}) all
diverge as $1/\sqrt{w}$ when $w \to 0$. Next, all these expressions vanish (for $n > 2$) as a power-law at the other edge of the support $[0,n]$, with an exponent dependent 
on  $n$. The case $n = 2$
is special: $P_{n=2}^{(2)}(w)$ diverges at both edges and has a minimum at $w = 1$, which signifies that in $2 \times 2$ Gaussian random matrices the two eigenvalues are
most probably very different from each other.
Further on, in Fig.~(\ref{fig0}) we plot these explicit forms together with more
lengthy expressions for $n = 7$ and $n = 12$. One observes that for $w < 1$ the distributions of arbitrary order are very close to the asymptotic result in Eq.~(\ref{sc}). The distibutions are multimodal indicating a set of probable and unprobable values of $w$, which mirrors certain structuring of the eigenvalues. 
As $n$ gets progressively
larger, the distributions become closer to the asymptotic result, Eq. ~(\ref{sc}), for any $w \in [0,4]$. Curiously enough, despite a rather complicated form of the polynomials in the second line of Eqs. (\ref{main}) and (\ref{main2}), they all show an appreciable variation with $w$  only for $w < 4$ and are indistinguishable from zero for larger values of $w$, despite the fact that formally  their support extends to larger than $4$ values of $w$.

For arbitrary $n$, an asymptotic behavior of $P_{n}^{(2)}(w)$ for $w \ll 1$ and
$w$ close to $ n$ can be readily deduced from Eq.~(\ref{main}). As we have already remarked, one finds that for $w \to 0$, the distribution
shows a generic singular behavior of the form:
\begin{equation}
P_{n}^{(2)}(w) \sim \frac{ \Gamma\left(n^2/2\right) \, _2F_1\left(-n+1,\frac{1}{2},2,2\right)}{\sqrt{\pi n} \, \Gamma\left((n^2-1)/2\right)} \, \frac{1}{\sqrt{w}} \,,
\end{equation}
where the amplitude
\begin{equation}
\frac{ \Gamma\left(n^2/2\right) \, _2F_1\left(-n+1,\frac{1}{2},2,2\right)}{\sqrt{\pi n} \, \Gamma\left((n^2-1)/2\right)} \to \frac{1}{\pi}
\end{equation}
when $n \to \infty$, in agreement with the general result in Eq.~(\ref{sc}). This implies, in turn, that fo a given random matrix a randomly chosen eigenvalue will most probably be much less than the 
arithmetic mean of all eigenvalues. 
Further on, on the opposite
extremity of the support, when $w$ is close to $n$, we have from Eq.~(\ref{main}) that
\begin{equation}
P_{n}^{(2)}(w) \sim \frac{2^{n-1}}{\sqrt{\pi} (n-1)!} \, \frac{\Gamma\left(n^2/2\right)}{n^{(n^2 - 2 n + 2)/2} \Gamma\left((n-1)^2/2\right)} \, \left(n - w\right)^{(n^2 - 2 n - 1)/2} \,,
\end{equation}
i.e., $P_{n}^{(2)}(w)$ attains a zero value as a power-law when $w \to n$ with an exponent which grows in proportion to $n^2$ when $n \to \infty$. This implies, in turn, that for $w$ sufficiently close to $n$ the value of $P_{n}^{(2)}(w)$ decays faster than exponentially with $n$.

Finally, we address the question 
how the Mar\v{c}enko-Pastur law in Eq. (\ref{sc}) can be  
derived from our Eq. (\ref{main}).  Below we briefly outline the steps involved in such a derivation. Note first that for $w < n$, one has 
\begin{equation}
\left (1 - \frac{w}{n}\right )^{(n^2-2 n - 1)/2} \to \exp\left (- \frac{n w}{2}\right ) \,,
\end{equation}
as $n \to \infty$, so that
\begin{equation}
\label{over}
\frac{\Gamma(n^2/2)}{\sqrt{\pi} \, \Gamma\left((n^2-1)/2\right) n^{3/2}} \, \frac{\left(1 -w/n\right)^{(n^2-2 n - 1)/2}}{ \sqrt{w}} \to \frac{\exp\left (- n w/2\right ) }{\sqrt{ 2 \pi n w}} \,.
\end{equation}
Further on, one has that, as $n \to \infty$, \cite{Abram}
\begin{equation}
P_{k}^{((n^2-3)/2-k,-1/2)} \left(2 \, \frac{w}{n} - 1\right) \to \frac{1}{4^k k!} \, H_{2 k}\left (\sqrt{\frac{n w}{2}}\right ) \,.
\end{equation}
Using next the integral representation of the Hermite polynomials
\begin{equation}
\fl
 H_{2 k}\left (\sqrt{\frac{n w}{2}}\right ) = \frac{\exp\left (n w/2\right ) }{\sqrt{\pi}} \, \left (\frac{2}{n w}\right )^{k+1/2} \, \int_{0}^{\infty} dy \, y^{2 k} \, \exp\left ( - \frac{y^2}{2 n w}\right ) \, \cos\left (y - \pi  k\right ) \,,
 \end{equation} 
 we can resummate the series in Eq. (\ref{main}) to find that, as $n \to \infty$, the sum in the latter equation converges to
 \begin{eqnarray}
 \label{int}
 \fl
 \sum_{k=0}^{n-1} {n \choose k + 1} \,
2^k \, \left(1 - \frac{w}{n}\right)^{n-k-1} \, P_{k}^{((n^2-3)/2-k,-1/2)} \left(2 \, \frac{w}{n} - 1\right) \to \nonumber\\ \to \sqrt{\frac{2}{\pi n w}}  \, \exp\left (\frac{n w}{2}\right ) \, \int_{0}^{\infty} dy \, \exp\left ( - \frac{y^2}{2 n w}\right ) \, \cos(y) \,  L_{n - 1}^{(1)}\left (\frac{y^2}{n w}\right ) \,. 
 \end{eqnarray}
Note next that as $n \to \infty$ \cite{Abram}
\begin{equation}
 L_{n - 1}^{(1)}\left (\frac{y^2}{n w}\right ) \to \frac{n \sqrt{w} J_{1}\left (2 y/\sqrt{w}\right ) }{y} \,,
 \end{equation} 
 so that the integral on the right-hand-side of Eq. (\ref{int}) converges to
 \begin{equation}
 \label{intint}
 \sqrt{\frac{2 n}{\pi}}  \, \exp\left (\frac{n w}{2}\right ) \, \int_{0}^{\infty} \frac{dy}{y} \, \exp\left ( - \frac{y^2}{2 n w}\right ) \, \cos(y) \,   J_{1}\left (\frac{2 y}{\sqrt{w}}\right ) \,,
 \end{equation}
 where $J_1(x)$ is the Bessel function.
 Noticing finally that 
 \begin{equation}
  \exp\left ( - \frac{y^2}{2 n w}\right ) \to 1
 \end{equation}
 as $n \to \infty$,  we can perform the integral in Eq. (\ref{intint}) in this limit, to get
 \begin{equation}
 \sqrt{\frac{2 n}{\pi}}  \, \exp\left (\frac{n w}{2}\right ) \, 
 \left\{
\begin{array}{rl}
\sqrt{1 - w/4} , \mbox{  for  } 0 < w < 4, \\
0 ,  \mbox{   for  } w > 4 .
\end{array}
\right.
\label{scc}
 \end{equation}
On combining the latter equation with Eq. (\ref{over}), we arrive at the Mar\v{c}enko-Pastur law in Eq. (\ref{sc}) .

\section{Conclusions}
\label{conclusion}

To recap, we analyzed the probability
distribution function $P_{n}^{( \beta)}(w)$
of the Schmidt-like random variable
$w = x_1^2/\left(\sum_{j=1}^n
    x^{2}_j/n\right)$, Eq. (\ref{def}),
    where $x_j$ are the eigenvalues of a given $n \times n$ $\beta$-Gaussian
    random matrix. 
    This variable, by definition, can be
considered as a measure
of how any individual eigenvalue deviates from
the arithmetic mean value of all eigenvalues of
a given random matrix, and its distribution is 
calculated with respect to the ensemble of such $\beta$-Gaussian
    random matrices. 
We showed that for arbitrary Dyson symmetry index $\beta$ in the asymptotic limit $n \to \infty$ the distribution  $P_{n}^{( \beta)}(w) $ converges to the  Mar\v{c}enko-Pastur form, i.e., is defined as $P_{n}^{( \beta)}(w) \sim \sqrt{(4 - w)/w}$ for $w \in [0,4]$ and equals zero outside of the support.
For Gaussian unitary ($\beta = 2$) ensembles 
we presented exact explicit expressions
for $P_{n}^{( \beta=2)}(w)$ valid for arbitrary $n$. We realised that, in general, $P_{n}^{( \beta=2)}(w)$ has a multimodal form indicating probable and unprovable values of $w$, which mirrors certain structuring of the eigenvalues.  We realised that the convergence to the Mar\v{c}enko-Pastur form is rather fast, so that already for $n = 12$ the exact result appears to be quite close to the asymptotic form.

\ack

The authors acknowledge helpful discussions with Oriol Bohigas and Satya N Majumdar.
This work is supported by the Brazilian agencies CNPq and FAPESP. GO is
partially supported by the ESF Research Network "Exploring the Physics
of Small Devices".

\end{document}